# Inversion Domain Boundary Induced Stacking and Bandstructure Diversity in Bilayer MoSe$_2$

*Jinhua Hong[†,‡,§], Cong Wang[†,§], Hongjun Liu[∥,∥∥,§], Xibiao Ren[‡], Jinglei Chen[∥],*

*Guanyong Wang[††], Jinfeng Jia[††], Maohai Xie[∥,\*], Chuanhong Jin[‡,\*], Wei Ji[†,\*],*

*Jun Yuan[‡,‡‡], Ze Zhang[‡]*

[†]Beijing Key Laboratory of Optoelectronic Functional Materials & Micro-Nano Devices, Department of Physics, Renmin University of China, Beijing 100872, China

[‡]State Key Laboratory of Silicon Materials, School of Materials Science and Engineering, Zhejiang University, Hangzhou, Zhejiang 310027, China.

[∥]Physics Department, The University of Hong Kong, Pokfulam Road, Hong Kong,

[††]Key Laboratory of Artificial Structures and Quantum Control (Ministry of Education), Department of Physics and Astronomy, Shanghai Jiaotong University, 800 Dongchuan Road, Shanghai 200240, China

[‡‡]Department of Physics, University of York, Heslington, York, YO10 5DD, United Kingdom

[∥∥]Institute of Functional Crystals, Tianjin University of Technology, Tianjin 300384, China.

§ These authors contributed equally to this work.

Correspondence and request for materials should be addressed to W.J. (wji@ruc.edu.cn), C.J. (chhjin@zju.edu.cn) or M.X. (mhxie@hku.hk).

Abstract:

Interlayer rotation and stacking were recently demonstrated as effective strategies for tuning physical properties of various two-dimensional materials. The latter strategy was mostly realized in *hetero*-structures with continuously varied stacking orders, which obscure the revelation of the intrinsic role of a certain stacking order in its physical properties. Here, we introduce inversion-domain-boundaries into molecular-beam-epitaxy grown $MoSe_2$ *homo*-bilayers, which induce uncommon fractional lattice translations to their surrounding domains, accounting for the observed diversity of large-area and uniform stacking sequences. Low-symmetry stacking orders were observed using scanning transmission electron microscopy and detailed geometries were identified by density functional theory. A linear relation was also revealed between interlayer distance and stacking energy. These stacking sequences yield various energy alignments between the valence states at the $\Gamma$ and K points of the Brillouin zone, showing stacking dependent bandgaps and valence band tail states in the measured scanning tunneling spectroscopy. These results may benefit the design of two-dimensional multilayers with manipulable stacking orders.

Van der Waals (vdW) epitaxy of two-dimensional (2D) layers has been demonstrated a marvelous route to build 2D nanostructures functionalized as transistors[1-6], photo detectors[7-11], light absorbers[6, 12], memories[13, 14], switchers[15] and other electronic and optoelectronic devices[16, 17]. Interlayer interaction plays a dominant role in determining physical properties of these nanostructures. An exceptional interlayer coupling mechanism, namely covalent-like quasi-bonding, leads to strongly layer-dependent evolution of electronic and vibrational properties, e.g. electronic bandgap, in black phosphorus[6, 18, 19] and $PtX_2$ (X=S or Se)[20-22]. Interlayer stacking order was predicted another degree of freedom to modify electronic structures of layered materials and recently demonstrated in twisted homo-[18, 23-25] and vdW hetero-bilayers[26-28]. Twisted homo-bilayers aside, hetero-bilayers involve electronic states from both different layers with a collective feature from diverse and continuously varied Moiré stacking orders, as a result of non-negligible lattice mismatch. It is, thus, of considerable importance to build a bilayer platform without interlayer lattice mismatch and rotation, e.g. a non-twisted homo-bilayer, to depict a more simplified but substantial physical picture on the correlation of stacking order and physical properties, e.g. electronic structures.

Molecular beam epitaxy (MBE) has recently been adopted to synthesize atomically thin transition metal dichalcogenides (TMDs)[21, 29-33]. With certain procedures, network-like inversion domain boundaries (IDBs) were introduced into $MoSe_2$ monolayers, behaving as metallic mid-gap states with signature for undergoing charge density wave transition at low temperature[29, 34]. In these monolayers, triangular domains are separated by IDBs forming domain-by-domain antiphases and fractional lattice translations. If two of these monolayers are stacked, randomly appeared IDBs in each layer should give rise to diverse stacking orders in a homo-bilayer; this provides a much improved platform for correlating stacking geometry with its electronic, optical or mechanical properties.

Here, we successfully grew a MoSe$_2$ *homo*-bilayer with randomly distributed IDBs through MBE, in which diverse unexpected low-symmetry stacking orders were discovered using aberration corrected transmission electron microscopy (AC-TEM)[35]. The details of these stacking orders were identified by comparing the experimental images with the simulated images based on the geometries revealed by density functional theory (DFT). In addition, DFT suggests stacking-dependent electronic structures, consistent with the domain-dependent spectra acquired using scanning tunneling spectroscopy (STS). Given the comparison of these results, we managed to build correlations between the observed geometries and measured electronic structures, and thus identified at least six low symmetric stacking orders that were not previously reported. A linear-scaling relation was established between interlayer distance and stacking stability, while exponential laws were found for the distance-dependent $C_Q$-$V_K$ or $C_Q$-$V_\Gamma$ gaps between valence (V) and conduction (C) bands. Here, Q, K and $\Gamma$ stand for three specific points in the Brillion zone, respectively. The competition of these two gaps leads to the band tail state observed in STS, as assessed to a stacking order with rather small interlayer distance. It is, to the best of our knowledge, the first time for the realization of large area, geometrically uniform and low symmetry stacking orders in vdW bilayers. Our work unveils the effects of ordered stacking on the electronic structures of TMD bilayers, which shed new light on tailoring the properties of 2D multilayers.

## Results

**Inversion domain boundaries in monolayer MoSe$_2$.** Figure 1a shows an atomically resolved annular dark field scanning transmission electron microscopy (ADF-STEM)[35] image of a MBE-grown MoSe$_2$ monolayer. The fast Fourier transform (FFT, Figure 1a inset) of the image shows unusual lines connecting those diffraction spots. This line-shaped feature is, most likely, a result of line defects in the monolayer, namely, inversion domain boundary (IDB), as denoted with blue ribbons in Figure 1b. These boundaries, densely embedded among adjacent MoSe$_2$ domains, are highly symmetric, atomically sharp, tri-atom wide, and coherent with the hexagonal lattice. Geometric phase analysis (GPA) of Figure 1a, as shown in Fig S1a-e, suggests that the breaking of lattice periodic symmetry is highly concentrated at the boundaries of these triangular domains. In Figure 1b, blue stripes highlight the domain boundaries among gold-colored continuous domains of monolayer MoSe$_2$.

A zoomed-in ADF-STEM image of the IDB was shown in the upper panel of Figure 1c. Brighter spots indicate Se$_2$ columns and darker ones for Mo atoms. An associated atomic model, illustrated with a 7×7 diamond-shape supercell (see Figure S2), was proposed for the boundary, as shown in Figure 1d. Both experiment and theory suggest that these triangular domains are terminated with Se$_2$ zigzag edges, in others words, two adjacent domains share the same line of Se$_2$ columns. The ADF image was simulated by QSTEM[36] (lower panel of Figure 1c) based on the fully relaxed atomic structure from DFT. Figure 1e shows the comparison of the ADF intensity line profiles along the long sides of the red and green rectangular stripes (marked in Figure 1c) of the experimental and simulated images. Both profiles slightly lack mirror symmetry at the boundary, which is a result of the unintentional residual aberration[37] (Supplementary Figure S3) of the STEM. This asymmetry was also observed by previous reports[30, 37] and does not obscure the structural determination of the IDBs. The good agreement of the comparison, together with the results from Lehtinen's work[30], convincingly supports the present atomic model of the $Se_2$-core boundary.

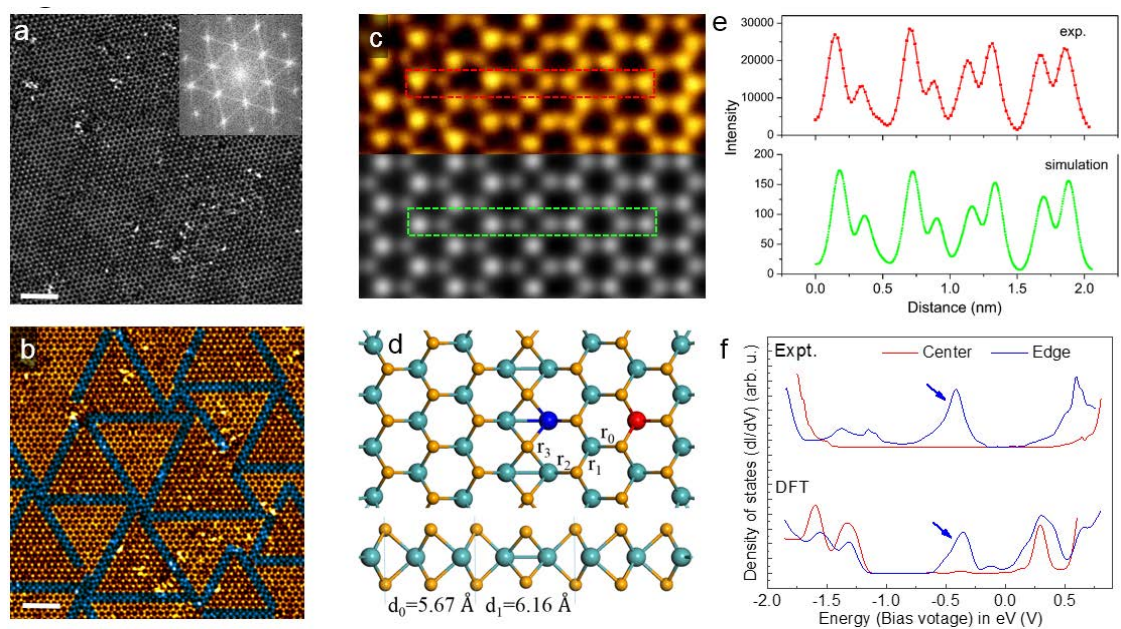

**Figure 1.** Inversion domain boundaries in MBE-grown monolayer $MoSe_2$. (**a**) High-resolution ADF-STEM image of monolayer $MoSe_2$. The inset FFT shows the quasi-periodicity of the ultra-narrow and long nanostructures. Scale bar: 2nm. (**b**) False colored domains and boundaries. These dense inversion domain boundaries connect with each other like wagon wheel. Scale bar: 2nm. (**c**) Experimental and simulated ADF images of the boundary. Scale bar: 0.5nm. (**d**) DFT relaxed atomic model of the boundary where orange balls represent Se atoms and cyan ones for Mo atoms. (**e**) ADF intensity profiles along the long sides of the rectangular stripes marked in **c**. (**f**) DFT calculated DOS and experimental STS spectra from the domain center and the boundary. The DOS data were acquired from the two Mo atoms marked in blue (boundary) and red (domain) balls shown in **d**. Here we focus on Mo atoms since S atoms only show negligible DOS around the pristine bandgap.

Our model was further verified by comparing measured scanning tunneling spectra (STS) (upper panel of Figure 1f) with theoretical density of states (DOS) (lower panel of Figure 1f) acquired at the boundary edge (blue curves) and domain center (red curves). The boundary-edge spectrum yields a mid-gap state at around -0.41 V and another two peaks at -1.8 V and 0.6 V. Our theory indicates the mid-bandgap state at -0.36 eV and other two peaks at -1.3 eV and 0.3 eV, highly consistent with those STS values. The domain-center spectrum shows a large bandgap slightly over 2.0 eV around the Fermi Level and the calculation unveils a bandgap of 1.5 eV, in accordance with the experiment in spite of the slightly underestimated bandgap by DFT.

**Diverse stacking orders in bilayer MoSe$_2$.** The IDB induces a mirror image domain at its other side. In addition, Figure 1d shows significant changes of Mo-Se bond lengths near the boundary, namely bond $r_2$ shortens from 2.54 ± 0.01 Å of the $r_0$ or $r_1$ value to 2.47 Å and bond $r_3$ elongates to 2.63 Å. These changes, together with newly formed Mo-Mo bonds, enlarges the horizontal Se-Se distance from 5.67 Å ($d_0$) to 6.16 Å ($d_1$). As a result, the boundary induces a lateral shift of 0.49 Å to the horizontal Se-Se period of 5.67 Å. This uncommon lateral translation, together with the mirror-image domains separated by the boundaries, inevitably bring about diverse stacking configurations[38], especially some low symmetric stacking orders, in a grown bilayer (as elucidated in the Supplementary word file). A key advance of this MBE MoSe$_2$ *homo*-bilayers lies in that the same stacking order could be, ideally, kept in a relative large domain area, different from the lattice-mismatch-induced Moiré stacking orders as very recently revealed in hetero-bilayers, e.g. MoS$_2$/WSe$_2$ bilayer[27].

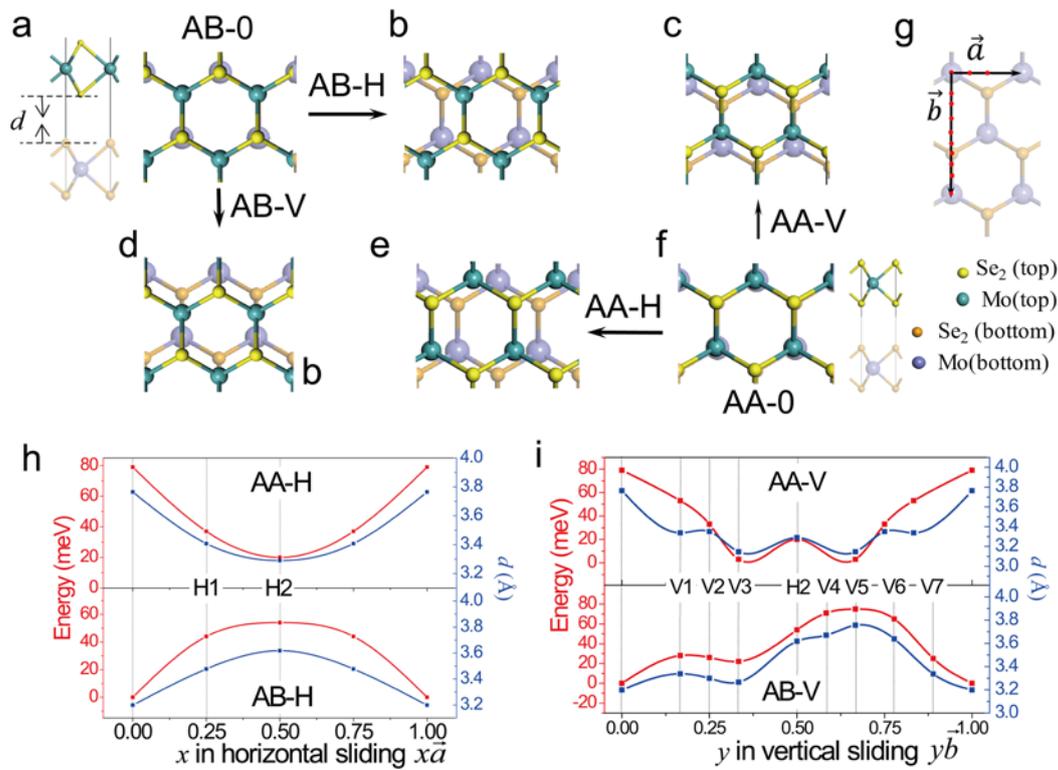

**Figure 2.** Schematic models and energetics of various stacking orders after in-plane translation. (a-f) Translational derivatives of the high-symmetry AA-0 and AB-0 stacking orders where the position of the bottom layer is fixed for better visualization. (g) Vectors $\vec{a}$ and $\vec{b}$ is defined as the period of the upper layer's horizontal/vertical sliding relative to the fixed bottom layer. The red points mark the intermediate positions of the relative sliding from the initial AA-0/AB-0. (h-i) Calculated total energy (red)

and interlayer distance $d$ (blue) as a function of horizontal or vertical sliding vectors, respectively, where the intermediate stacking orders are labeled.

The lattice sliding caused by the network-like IDBs should, in principle, lead to infinite numbers of stacking orders assuming ridge $MoSe_2$ layers. The actual number is, however, limited by the competition between the energy used to wrinkle the layer and the interlayer attraction between two laterally 'shifted' layers. To cover the possibility of stacking orders as complete as possible, we chose two high symmetry stacking orders, i.e. AB-0 and AA-0, as initial configurations. Structure AB-0 (Figure 2a) is the normal 2H stacking order found in natural $MoSe_2$ crystals where the two layers have inversion symmetry. Structure AA-0 (Figure 2f) is, however, the most unstable one among all considered stacking orders, in which both $MoSe_2$ layers have exact mirror symmetry to a mirror plane between them. Sixteen stacking orders were thus constructed by sliding the top layer along four pathways initialized from AB-0 and AA-0 as illustrated in Figure 2a-f. Figure 2g indicates all exact positions of the initial, final and intermediate configurations in red dots for both horizontal ($\vec{a}$) and vertical ($\vec{b}$) paths. Some of these configurations cannot stably hold their initial stacking positions and transform into other stacking orders. We, therefore, kept their relatively lateral positons and obtained their optimized interlayer distances and total energies.

Figures 2h and 2i plot the distance-energy relation of these 16 considered configurations. Detailed geometries and exact values of the distances and energies are available in Supplementary Figure S4 and Table S1. Interlayer distance $d$ varies from 3.15 Å to 3.76 Å and the energy differs by up to 79 meV with a nearly linear dependence on the interlayer distance (See Supplementary Figure S5). This dependence, more comprehensive than the qualitative relation found in twisted $MoS_2$[23], is a result of the subtle balance between interlayer vdW attractions and Pauli/Coulomb repulsions, especially the repulsion between interlayer Se $p_z$ orbitals. For example, Mo and Se atoms of the top layer are exactly over those corresponding atoms of the bottom layer in configuration AA-0 (Figure 2f), leading to the strongest repulsion between the orbitals of interlayer Se atoms and thus the largest interlayer distance $d$ and the highest total energy (Figure 2h upper panel). For AB-0 bilayers (Figure 2a), Se atoms of the top layer sit over the hollow sites of Se triangles of the bottom layer, giving rise to a substantially shortened distance and much lowered total energy (Figure 2h lower panel). An exception was found for AA-V1 (Figure 2i) that its energy is roughly 25 meV higher than AB-V1 but they share nearly identical interlayer distances. These differences of interlayer distance were believed observable as various apparent heights in STM measurements being discussed in Supplementary Figure S8. All these results also imply that different stacking orders may affect electronic structures as we elucidated later.

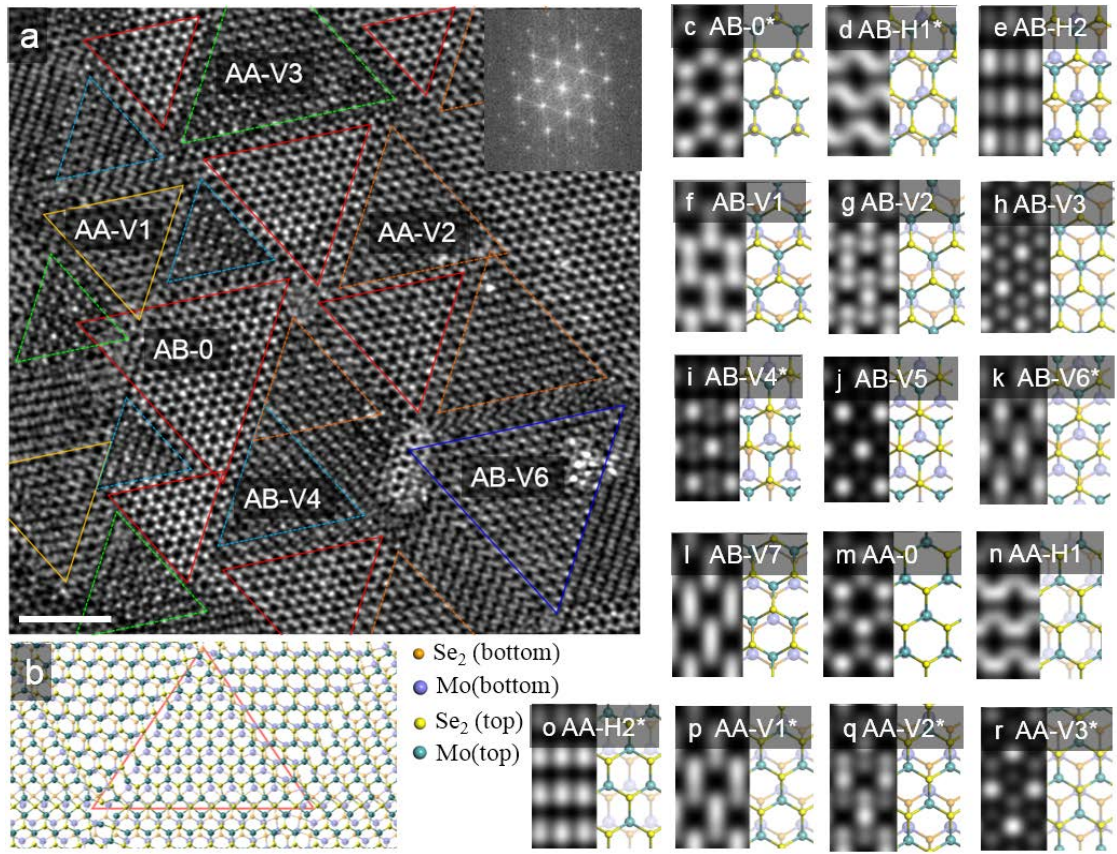

**Figure 3.** Diverse atomic structure of bilayer domains. (**a**) Experimental ADF image of a typically continuous and uniform bilayer $MoSe_2$. The triangles in the same color indicate these domains are in the same stacking order. Scale bar: 2nm. (**b**) Atomic model of diverse bilayer domains with stacking orders induced by the intrinsic IDBs in monolayer. (**c-r**) Simulated ADF images from different bilayer stacking orders. The atomic structures of each domain in **a** are assigned by the comparison of experimental and simulated ADF images. Eight stacking orders (marked with symbol *) were observable in the experimental ADF images, and six of them can be found in **a.**

Figure 3a shows an atomically resolved ADF image of bilayer MBE-$MoSe_2$, in which different triangular domains are observable, with diverse topographies resulted from the vast IDBs. As previously reported[38], IDBs exist in both $MoSe_2$ layers. We show a sketch in Figure 3b to illustrate the origin of the diversity, which is comprised of two MoSe$_2$ layers with IDBs in each of them, yielding naturally diverse ordered stacking domains. More details are available in the Supplementary word file. The FFT pattern (Figure 3a inset) shows no rotation angle between the upper and bottom layers. This result, together with the non-uniform and non-periodic domains, rules out the possibility of these patterns being Moiré pattern, but supports the fact that they are relevant with various stacking orders. Simulated ADF images of 16 stacking orders and their corresponding atomic structures are shown in Figure 3c to 3r. By comparing them with the experimental images, eight models (marked with symbol *) among the total 16 were assessed experimentally observable, while six of them (shown in Figure 3c, 3i, 3k, 3p, 3q and 3r) were available in the area shown in Figure 3a and the other two were shown in Supplementary Figure S6.

The most stable stacking order AB-0 (Figure 3c) was found the most common triangle domain in our ADF images. The second most stable one, AA-V3 (Figure 3r), is also frequently found, which is consistent with the order of thermal stability of these stacking orders. These two most stable stacking orders correspond to those bilayer configurations in the well-known 2H and 3R phases, respectively. As we discussed earlier, AA-V1 (Fig. 3p) is a fairly less stable, low symmetrical and fractional translation induced stacking structure. It was, however, also found in our experimental images (Figure 3a), owing to the confinement of the IDBs. In addition to AA-V1, other four configurations, shown in Fig. 3d, 3i, 3k and 3q, were theoretically found unstable in pristine bilayers but were experimentally observed in STEM images. It is, to the best of our knowledge, the first time that uniform, low symmetrical and fractional translation induced stacking orders are prepared in a *large domain* of *homogeneous* bilayer TMDs, essentially different from the various stacking orders recently observed in Moiré patterns of hetero-bilayer TMDs[26, 27].

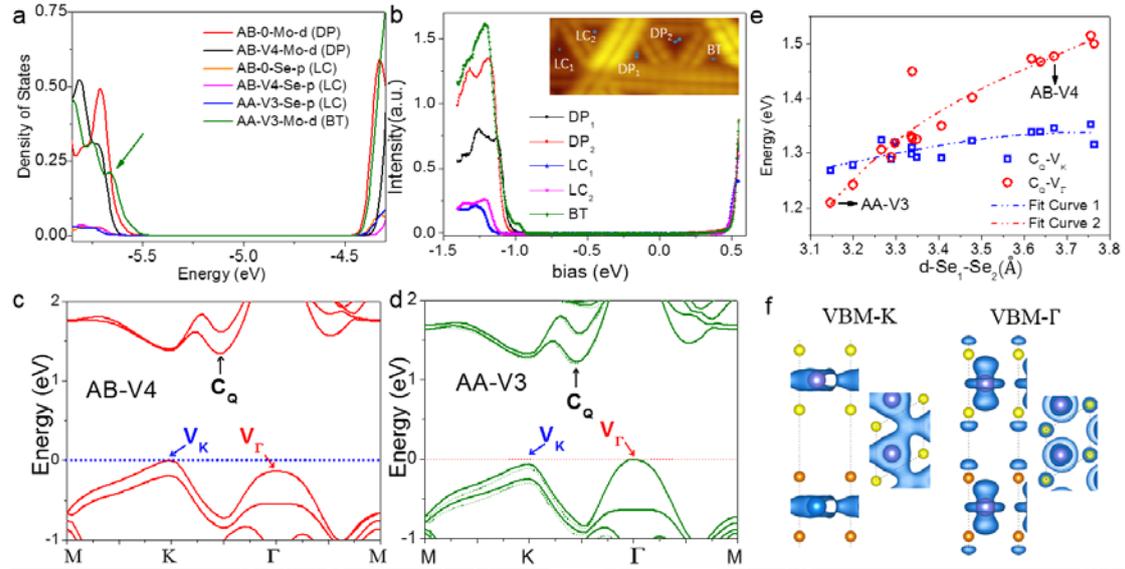

**Figure 4.** Distinctive electronic structures of the diverse bilayer domains. (**a**) DFT calculated LDOS of several typical bilayer stacking structures. The valence band edge is dependent on the stacking order. (**b**) Experimental STS spectra measured at different domains. The different band tail states should arise from the diverse stacking orders of bilayer domains. The inset is a STM image of the corresponding domains taken at a sample bias voltage of -0.59 V and set point current of 96 pA. (**c-d**) The band structures of pristine bilayer in stacking orders AB-V4 and AA-V3. (**e**) Calculated $C_Q$-$V_K$ (black) and $C_Q$-$V_\Gamma$ (red) gaps as a function of interlayer distance for different stacking orders. (**f**) Side- and top views of the partial charge densities (wavefunction distributions) of the states at $V_K$ and $V_\Gamma$ of AA-0. The choice of AA-0 is for clarity only and the shapes of wavefunction for both states are essentially the same for all stacking configurations.

**Electronic structures of the diverse bilayer domains.** Various low-symmetry stacking orders in large domains, a unique feature of this work, may induce novel electronic structures that could be feasibly used in multilayer or heterostructure devices for potential applications. Visualized wavefunctions indicate that most of the valence and conduction states are comprised of Mo *d*-states and a small portion of them is contributed from Se *p*-states. We thus plot the theoretically local density of states (LDOS) of Mo *d*- and Se *p*-states of three representative stacking orders, namely AB-0, AB-V4 and AA-V3 in Figure 4a. Here, AA-V3 and AB-0 are the most stable and experimentally observable AA and AB stacking configurations, respectively, while AB-V4 is unstable in pristine bilayers but could be stabilized by IDBs in MBE-grown bilayers. The intensity of LDOSs of Mo *d*-states is substantially larger than that of Se *p*-states. The energetic position of the conduction band edge of either Mo *d*- or Se *p*-states is nearly unchanged regardless of stacking orders except slight shifts within 70 meV, however, that of the valence band edge varies by at least 110 meV, as noted by the arrow in Figure 4a. There are two categories of Mo valence band edges, namely a normal-appearance one (red curve and black curve in Figure 4a) and the other one with obvious band tail state (olive curves). The LDOS of configuration AB-V4 is a representative normal-appearance one. It has a bandgap of roughly 1.21 eV and the pronounced peak of valence band splits into two peaks with a 90 meV separation. We thus denote this category as ``Double-Peak (DP)''. For AA-V3 (olive), a band tail is explicitly observable, which reduces the band gap from 1.21 eV to 1.03 eV. ``Band-Tail (BT)'' is the tag of this category. In terms of the LDOSs of Se *p*-states, their valence band edges are even energetically lower than those of Mo *d*-states, leading to an "apparent" gap of 1.25 eV. The low intensity of this category gives rise to smaller conductance in STS measurements. We thus call this category ``Low-Conductance (LC)''.

All these features found in those LDOS plots were well obtained in STS experiments, as shown in Figure 4b. Those STS spectra are highly reproducible and exhibit consistent features within one domain, as show in Supplementary Figure S7. The measured bandgaps are 1.49 eV, 1.40 eV and 1.57 eV, respectively, comparable to those theoretical values. It is fairly challenging to identify the local stacking configuration of each domain solely based on STM images, as we illustrated in Supplementary Figure S9. It also remains difficult even with the inputs of STS spectra since a few stacking configurations share the same feature of STS spectra. However, we do be able to classify them into the three categories found in the LDOS calculations. The olive spectra showing a band tail feature in Fig. 4b was assigned to category BT, in which AA-V3 is a representative configuration, consistent with the fact that the theoretically stable stacking AA-V3 was frequently obtained in experimental ADF images. Spectra $DP_1$ and $DP_2$ showing double-peak features were thus assigned into category DP and spectra $LC_1$ and $LC_2$ correspond to category LC.

The appearance of LDOSs depends on stacking orders. We, therefore, plot the band structures of AB-V4 and AA-V3, two novel stacking configurations, in Figure 4c and 4d, respectively. It shows that the double-peak feature of the valence band of AB-V4 is originated from the VBM at K and the 130-meV-lower valence state at Γ. In AA-

V3, the VBM was, however, found at $\Gamma$. This flatter band is nearly degenerated with the 59-meV-lower valence state at K. This band alignment results in the band tail state. It would be interesting to unveil the correlation between the VBM location and stacking order. Figure 4e shows the K-Q (blue) and $\Gamma$-Q (red) gaps as a function of the interlayer distance for all considered stacking orders. Both gaps increase in a nearly exponential manner with respect to interlayer distance $d$ while the $\Gamma$-Q gap goes faster than the K-Q gap. The VB is primarily comprised of Mo $d_{z2}$ and Se-$p_z$ orbitals at the $\Gamma$ point while it is mainly confined in the Mo plane at the K point (Figure 4f). The energy level of VB at $\Gamma$, therefore, changes more speedily than that at K when the interlayer distance varies. The competition of VB states at $\Gamma$ and K results in the VBM locating at the $\Gamma$ point for stacking orders with smaller interlayer distances, e.g. AB-0 and AA-V3. In stacking orders with the smallest interlayer distance, i.e. AA-V3, the most strongly overlapped Se $p_z$ orbitals from both layers give rise to higher energy levels owing to Coulomb repulsion; this is thus observed as the band tail state, a fingerprint for smaller-interlayer-distance stacking orders, in STS measurements.

**Discussion**

One of the key advances of this work lies in the prediction and assessment of the homogeneous bilayer domains with diverse certain stacking orders, which is essentially resulted from the network-like IDBs. The derivative lattice sliding caused by IDBs, therefore, participates in the competition between the layer-layer attraction (over 200 meV) and orbital-orbital repulsion induced by deviation from the most favored configuration (up to 80 meV). In light of this, apart from the stable AB-0 and AA-V3 where interlayer attraction dominates, those configurations predicted not even meta-stable in pristine bilayer, e.g. AB-V4 and AA-V2 are likely accessible in our bilayer. We managed to observe eight stacking orders in our AC-TEM images. All of them were identified to be among the totally 16 DFT predicted ones. We thus expect that the rest configuration could be, most likely, obtained by finely tuning the density of IDBs as we infer the length of the lattice sliding is modifiable by the density of IDBs. The established linear relation between the interlayer distance and the difference of stacking energy suggests a fitted slope of 0.12 eV/Å. It was also found that different stacking orders do not appreciably change the conduction band, but gives rise to a competitive energy alignment of valence states at both the $\Gamma$ and K points of the Brillouin Zone. These observed stacking orders yield a plenty of relative positions in energy for the valence states at $\Gamma$ and K, showing stacking dependent bandgaps and valence band tail states in the measured STS. This work illustrates a successful demonstration for investigating of the stacking-bandgap diversity in 2D layered materials. In addition, it extends the family of uniform large-area novel stacking orders in bilayer TMDs and develops the knowledge of stacking order in modifying their electronic structures. All of these facts may benefit in band engineering of 2D electronics.

**Methods**

**Sample preparation and STM/STS.** Monolayer and bilayer $MoSe_2$ samples were prepared through molecular beam epitaxy on HOPG substrate at 450 ℃ in an Omicron UHV system. The fluxes of Mo and Se were generated from an e-beam evaporator and a Knudsen cell respectively and the film deposition was carried out under a Se-rich environment with a flux ratio as high as 15 between Se and Mo. Freshly cleaved substrate was degassed overnight at 550 ℃ in vacuum before sample growth. Reflection high-energy electron diffraction (RHEED) was employed for in situ surface analysis. STM measurements were carried out at 77 K using the constant current mode in a separated low-temperature Unisoku STM system. Before being taken out from the vacuum system, the sample surface was capped by an amorphous Se layer deposited at the room temperature, which was desorbed by annealing prior to the STM experiments. The latter was reflected by both the recovery of the streaky RHEED pattern and the clean and flat surface morphology revealed by STM examinations. Differential conductance spectra were taken at 77 K using a lock-in amplifier with a modulation voltage 15 mV and frequency 985 Hz. Each STS curve shown represented an average of 50 measurements at the same positions.

**TEM characterization and image simulation.** Atomically resolved ADF-STEM imaging was conducted inside an aberration corrected TEM (FEI Titan ChemiSTEM) at 200kV. A probe current 60 pA was used for the ADF imaging (with a detector acceptance angle β~50-200 mrad) to avoid beam irradiation damage. The convergence angle (α) of the incident electron beam was set to 21 mrad. The probe corrector help lower down the aberration Cs to 2 μm and atomic resolution ADF -STEM imaging is accessible under such experimental conditions. ADF-STEM image simulations were done with computer package QSTEM[36] under the same parameter settings as the experimental conditions such as Cs , α, β besides the probe size ~1.0Å, and residual astigmatism $A_2$ was set to 100 nm to match the experimental imaging of the inversion domain boundaries.

**DFT calculations.**

Density functional theory calculations were performed using the generalized gradient approximation[39] for the exchange-correlation potential, the projector augmented wave method[40, 41] and a plane-wave basis set as implemented in the Vienna ab-initio simulation package (VASP)[42]. Van der Waals interactions were considered at the vdW-DF level[43, 44] with the optB86b[45] exchange functional, which achieves accurate results in calculating structural properties of two dimensional materials[19, 20, 22, 46, 47]. A 7×7 supercell was adopted to model the IDBs and the domain areas among IDBs in $MoSe_2$ monolayer. Sixteen kinds of bilayer stacking configurations, modeled by 1×1 unit cell, were investigated to find out the variant properties deduced by the structure discontinuity. Kinetic energy cut-offs of 400 eV and 700 eV were adopted for the calculations of $MoSe_2$ monolayer with IDBs and the variant stacking orders in bilayer $MoSe_2$, respectively. Two $k$-meshes of $4 \times 4 \times 1$ and $13 \times 13 \times 1$ were used to calculate the first Brillouin zone of the 7×7 supercell and the 1×1 unit cell, respectively. The lattice parameters and cell volumes of all the configurations were fully optimized and all atoms were allowed to relax until the residual force per atom was less than 0.001 eV $Å^{-1}$.

## ASSOCIATED CONTENT

**Supporting Information Available:** Figures S1-9. Structure and strain analysis of inversion domain boundaries. Quantitative ADF image simulation of monolayer $MoSe_2$ and bilayer stacking configurations. Different STS spectra from within one domain in bilayer $MoSe_2$. This information is available free of charge via the Internet at http://pubs.acs.org

## AUTHOR INFORMATION

**Corresponding Authors**

* Correspondence and request for materials should be addressed to W.J. (wji@ruc.edu.cn), C.J. (chhjin@zju.edu.cn) or M.X. (mhxie@hku.hk).

**Author contributions**

J.H., C.W and H.L. contributed equally to this work. C.J., M.X. and W.J. conceived this research. J.H. and C.J. performed AC-TEM measurement and image simulation; H.L.,

J.C. and M.X. grown the sample and measured the STS data with help from G. W. and J.J.; C.W. and W.J. did all theoretical calculations. C.W., J.H, H.L., M.X., C.J. and W.J. wrote the manuscript and all authors comment on it.
**Notes**
The authors declare no competing financial interests.

# ACKNOWLEDGEMENTS

This work was financially supported by the National Science Foundation of China under grant Nos. 51772265, 11274380, 91433103, 51472215, 11622437, 61674171 , 61721005 and 51222202, the National Basic Research Program of China under grant Nos. 2014CB932500 and No. 2015CB921004, the 111 project (No. B16042), the Fundamental Research Funds for the Central Universities under grant Nos. 16XNLQ01 (RUC) and the State Key Laboratory of Clean Energy Utilization. M.X. acknowledges the support by a Collaborative Research Fund (HKU9/CRF/13G) of the Research Grant Council, Hong Kong Special Administrative Region. J.Y. acknowledges the EPSRC (UK) funding EP/G070474/1 and supports from Pao Yu-Kong International Foundation for a Chair Professorship in ZJU. This work made use of the resources of the Center of Electron Microscopy of Zhejiang University. Calculations were performed at the Physics Laboratory for High-Performance Computing of Renmin University of China and at the Shanghai Supercomputer Center.

Author ORCID information：J. H.: 0000-0002-6406-1780, C.W.: 0000-0002-5297-9586, H.L.：0000-0002-6656-2669, M.X.: 0000-0002-5017-3810, C. J.: 0000-0001-8845-5664, W.J.: 0000-0001-5249-6624.

List of abbreviations: Van der Waals (vdW), two-dimensional (2D), Molecular beam epitaxy (MBE), transition metal dichalcogenides (TMDs), inversion domain boundaries (IDBs), aberration corrected transmission electron microscopy (AC-TEM), density functional theory (DFT), scanning tunneling spectroscopy (STS), atomically resolved annular dark field scanning transmission electron microscopy (ADF-STEM), fast Fourier transform (FFT), Geometric phase analysis (GPA), Reflection high-energy electron diffraction (RHEED), Quantitative TEM/STEM Simulations (QSTEM), local density of states (LDOS), 'band-tail' (BT), 'double-peak' (DP), 'relative low conductance' (LC); (UHV) ultra-high vacuum; (HOPG) highly oriented pyrolytic graphite.

# Inversion Domain Boundary Induced Stacking and Bandstructure Diversity in Bilayer MoSe$_2$

*Jinhua Hong*[†, ‡, §]*, Cong Wang*[†, §]*, Hongjun Liu*[∥, #, §]*, Xibiao Ren*[‡]*, Jinglei Chen*[∥]*,*

*Guanyong Wang*[††]*, Jinfeng Jia*[††]*, Maohai Xie*[∥, \*]*, Chuanhong Jin*[‡, \*]*, Wei Ji*[†, \*]*,*

*Jun Yuan*[‡, ⊥]*, Ze Zhang*[‡]

[†]Beijing Key Laboratory of Optoelectronic Functional Materials & Micro-Nano Devices, Department of Physics, Renmin University of China, Beijing 100872, China

[‡]State Key Laboratory of Silicon Materials, School of Materials Science and Engineering, Zhejiang University, Hangzhou, Zhejiang 310027, China.

[∥]Physics Department, The University of Hong Kong, Pokfulam Road, Hong Kong,

[††]Key Laboratory of Artificial Structures and Quantum Control (Ministry of Education), Department of Physics and Astronomy, Shanghai Jiaotong University, 800 Dongchuan Road, Shanghai 200240, China

[⊥]Department of Physics, University of York, Heslington, York, YO10 5DD, United Kingdom

[#]Institute of Functional Crystals, Tianjin University of Technology, Tianjin 300384, China.

§ These authors contributed equally to this work.

Correspondence and request for materials should be addressed to W.J. (wji@ruc.edu.cn), C.J. (chhjin@zju.edu.cn) or M.X. (mhxie@hku.hk).

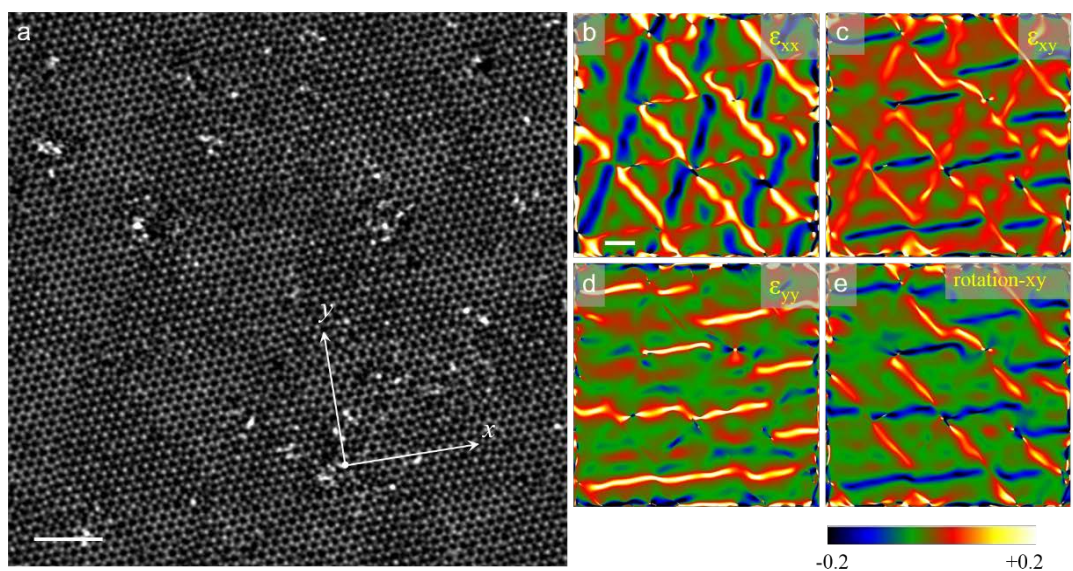

**Supplementary Figure S1. Inversion domain boundaries in MBE monolayer MoSe$_2$.** (a) High-resolution ADF-STEM image of monolayer MoSe$_2$. Scale bar: 2 nm. The x and y directions were defined by the white axes, representing the zigzag and armchair directions of the MoSe$_2$. (b-e) GPA analysis of the lattice strain $\varepsilon_{xx}$, $\varepsilon_{xy}$, $\varepsilon_{yy}$ and lattice rotation-*xy* to reveal the lattice deformation or discontinuity. The breaking of lattice periodic symmetry is found to be highly concentrated within the boundaries of the triangular domains.

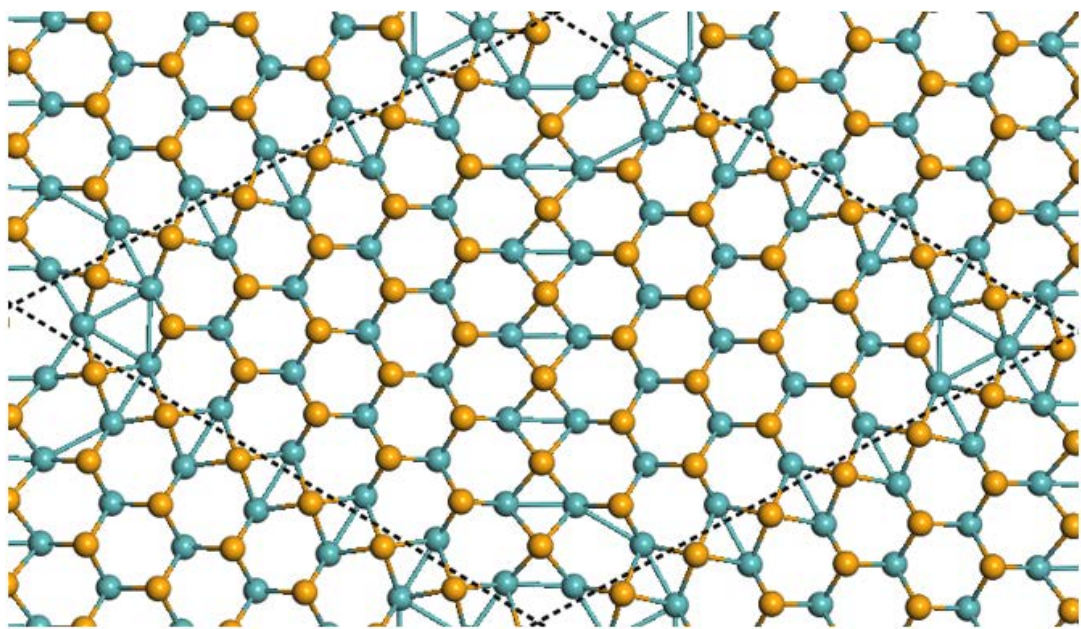

**Supplementary Figure S2. Top view of monolayer MoSe$_2$ with domain boundaries as modeled in a 7×7 diamond-shape supercell.** As model with same kind of boundaries is unachievable without threefold rotational symmetry in the MoSe$_2$ hexagonal lattice, the ribbon model is not suitable. To confirm the atomic structure of inversion domain boundaries in monolayer MoSe$_2$, a 7×7 diamond-shape supercell model was proposed. Further simulation and theoretically calculated density of states was based on this model.

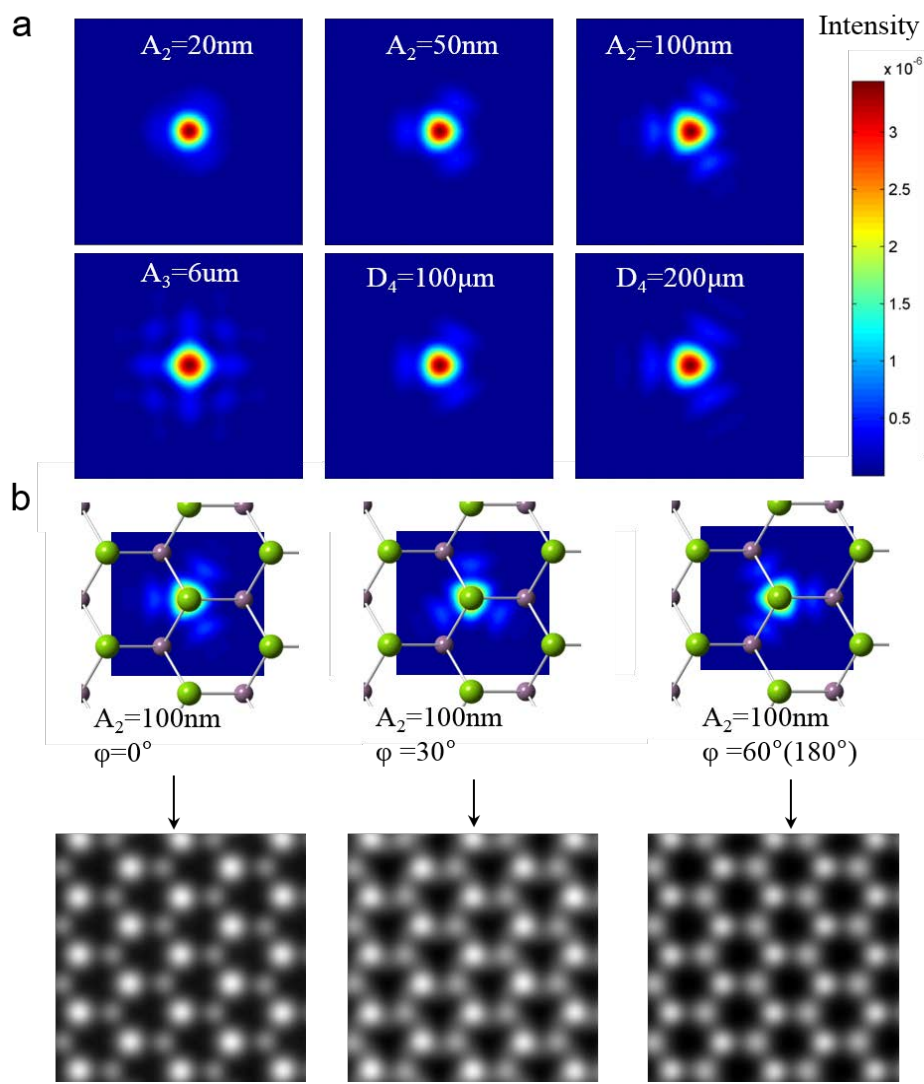

**Supplementary Figure S3. ADF-STEM imaging simulation of monolayer MoSe$_2$.** (a) The spatial distribution of the intensity of the focused electron probe in STEM imaging with aberrations such as 3-fold astigmatism A$_2$, 4-fold astigmatism A$_3$ and three lobe aberration D$_4$. These aberrations in the electron probe will make the beam deviate from the standard Gaussian intensity distribution, and turn into extended distribution with symmetric 'tails'. The color bar shows the scale of the intensity of the beam. (b) Simulated ADF-STEM images under residual aberrations. The ADF-STEM imaging intensity follows I$_{ADF}$=I$_{probe}$(χ)⊗U$_{lattice}$, which is a convolution of the electron probe I$_{probe}$ (aberration function χ is caused by unintentional residual aberration A$_2$, A$_3$, and relative phase angle φ) and the periodic lattice potential field U$_{lattice}$ (object function) of the crystal sample. For A$_2$-aberrated beam with different phase angles, the atomic model of MoSe$_2$ monolayer is fixed to show the effect of aberrations on the ADF imaging. The asymmetric ADF intensity at both sides of the boundary in Figure1 is, as revealed by detailed ADF-STEM imaging simulations, a result of the unintentional residual aberration.

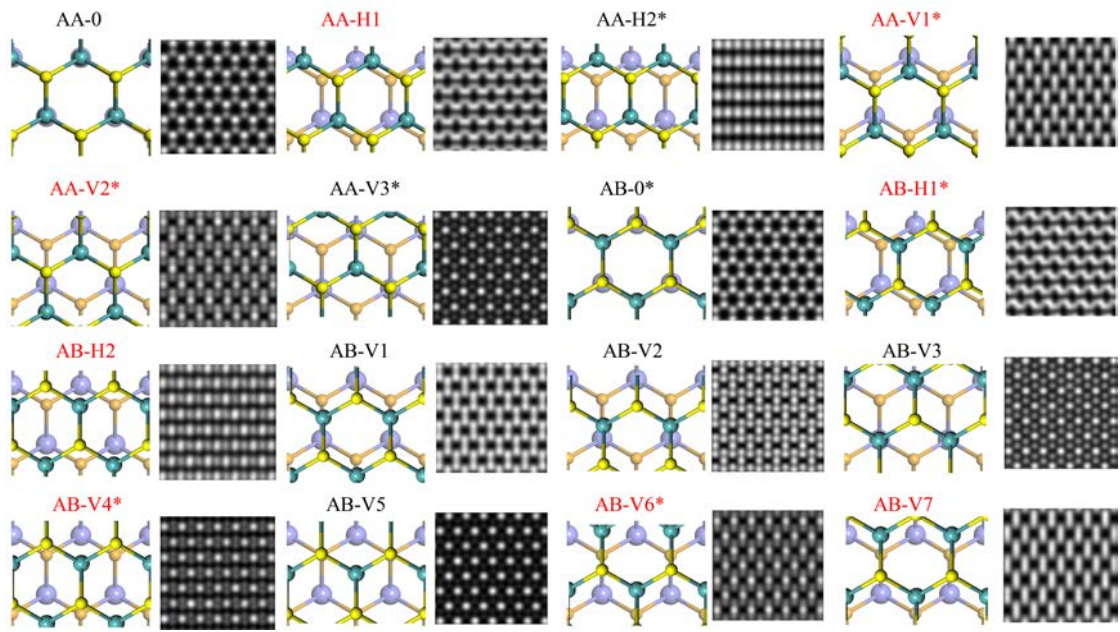

**Supplementary Figure S4. Top view structure and ADF-STEM simulation images of the calculated stacking orders of bilayer MoSe$_2$.** Stacking orders were constructed by sliding the top layer along four pathways initialized from the two configurations AA-0 and AB-0. The stacking orders with their names highlighted in red were theoretically found unstable in pristine MoSe$_2$ bilayers. They cannot hold their initial stacking positions and transform into other stacking orders after relaxation in pristine bilayers. Those stacking orders, marked with symbol *, were observable in experimental ADF-STEM images, including several unstable stacking orders in pristine bilayers, such as AA-V1 and AB-V4. In MBE-grown bilayers, they were stabilized by the confinement of the IDBs and thus can be experimentally observed.

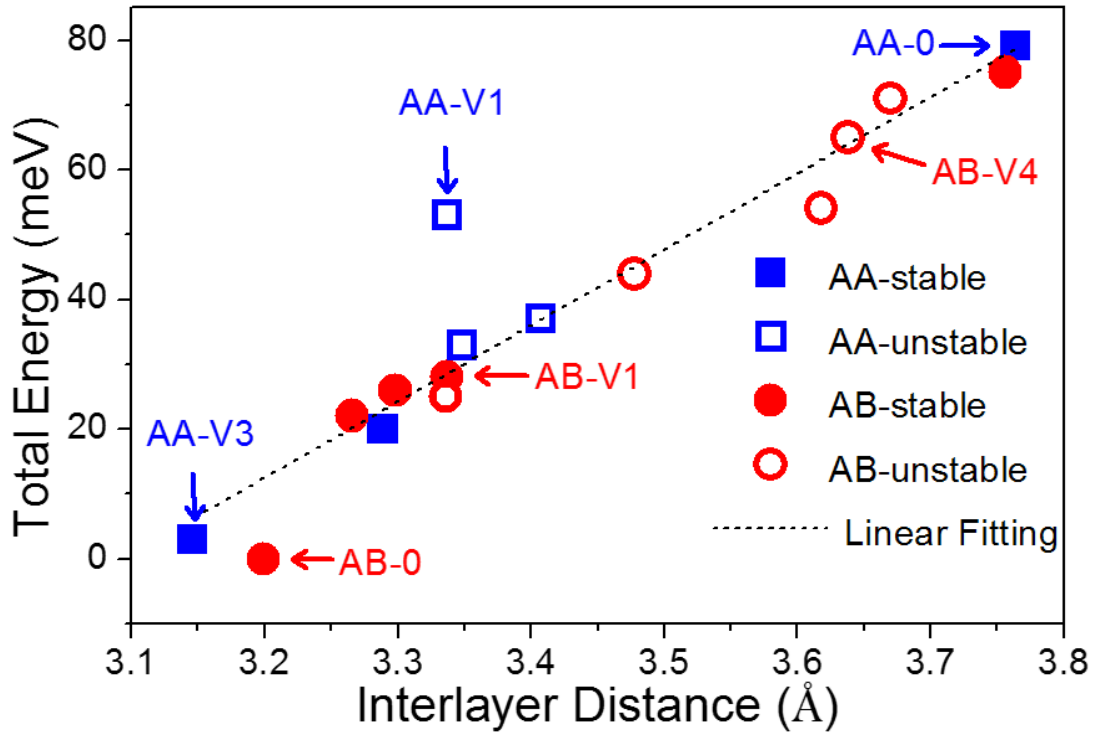

**Supplementary Figure S5. The calculated relative energy as a function of the interlayer distance $d$ for different stacking orders.** A linear relation was revealed between interlayer distance and stacking energy, which suggests a fitted slope of 0.12 eV/Å. The only exception is stacking configuration AA-V1. Its interlayer distance is close to that of AB-V1, however, its energy is 25 meV higher than that of AB-V1. We attribute the higher energy to the stronger repulsion between interlayer Se-$pz$ orbitals in AA-V1. As for AB-V1, top-layer Se atoms reside at the bridge site of two bottom-layer Se atoms, substantially lowering the repulsion energy. Stacking order AB-V4 is a typical unstable configuration with large interlayer distance, high relative energy and low symmetry. With the existence of IDBs, AB-V4 can be formed in MBE bilayer MoSe$_2$, whose statistical count of 7 is only less than that of most stable AB-0.

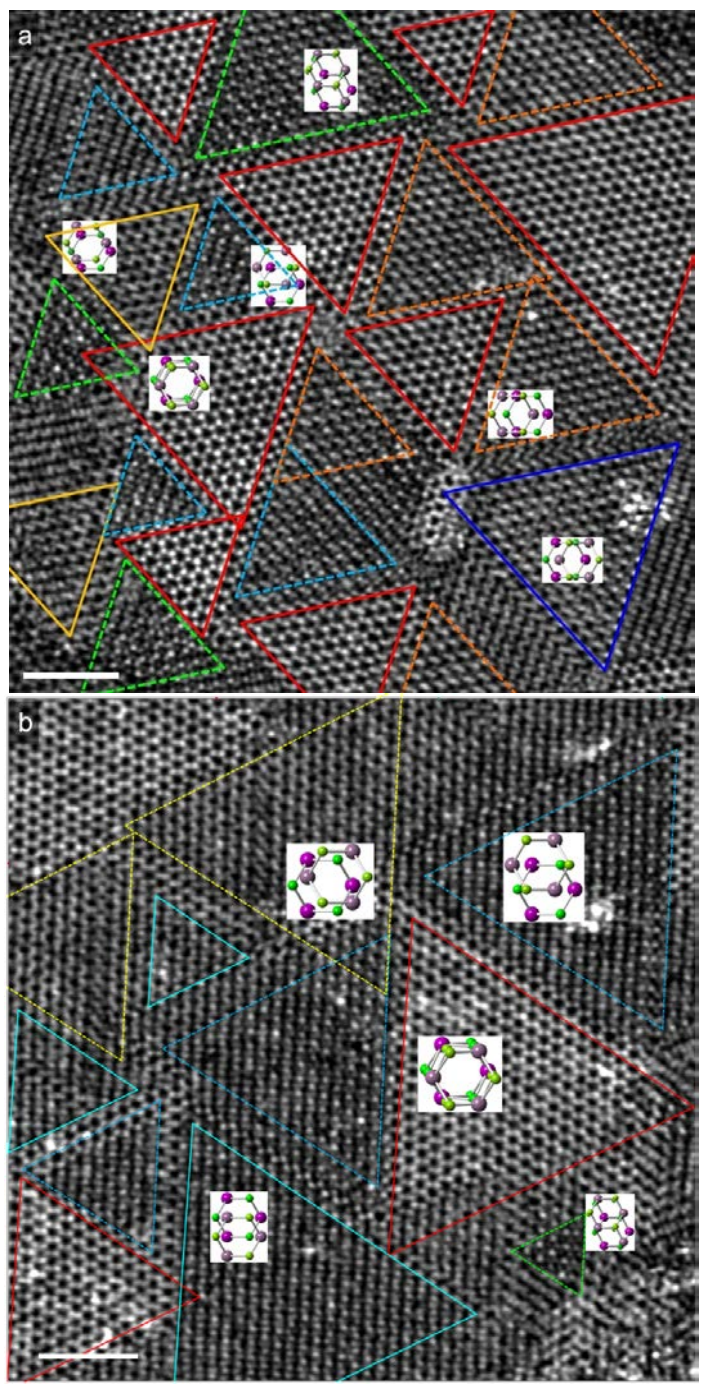

**Supplementary Figure S6. Experimental ADF images associated with their assigned atomic models of a typical continuous and uniform bilayer MoSe$_2$.** The triangles in the same color indicate the same stacking orders. Scale bars: 2nm. Eight stacking orders can be identified by comparing the simulated images with experimental observations. Configuration AB-0, the most stable stacking order theoretically revealed, corresponds to the most commonly observed domains in experimental statistics, as labeled with red triangles. Two configurations not available in Fig. 3a were shown in (b) marked with cyan (AA-H2) and light yellow (AB-H1) triangles.

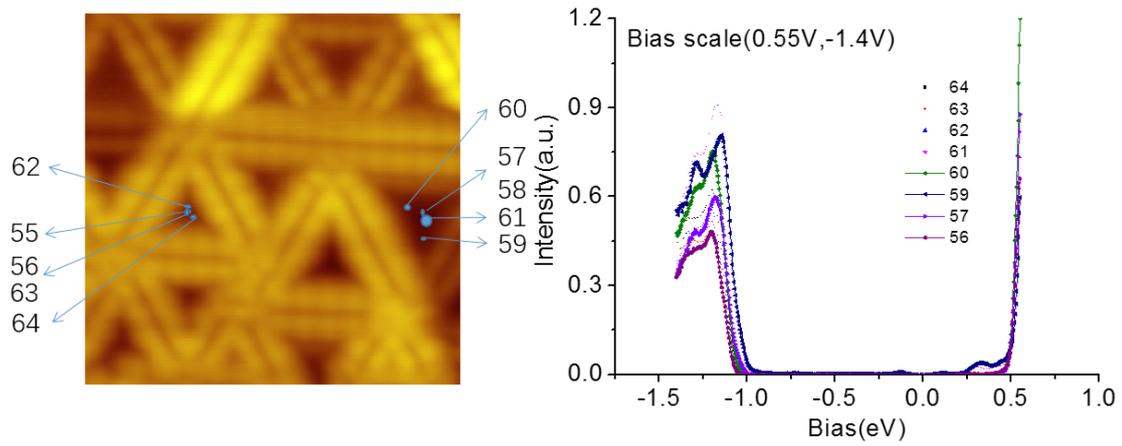

**Supplementary Figure S7. Experimental STS spectra of defect-free area inside triangular domain.** The experimental spectra are highly reproducible and show little change within one triangular domain, hence the stacking orders are responsible for the diversity of spectra.

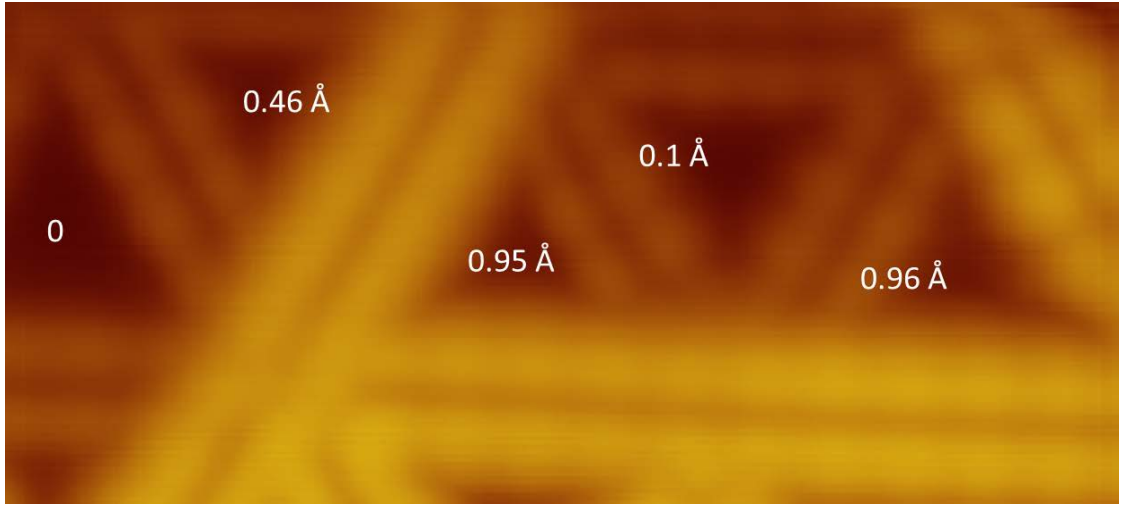

**Supplementary Figure S8. Apparent heights at the centers of the triangle domains in the STM image measured by line profiling.** The height is offset with respect to that of the very left domain marked by '0'. We measured the relative height of each domain (reference zero) and labeled their values in the figure. In certain cases, we even found apparent height difference larger than that enumerated in table S1. We noticed a tendency that the smaller the triangular domain size, the brighter the contrast. Thus, we infer the high intensities of the IDB defects in the image may have affected the contrast of the enclosed domains, which partially reduces the quantitative accuracy of height measurements.

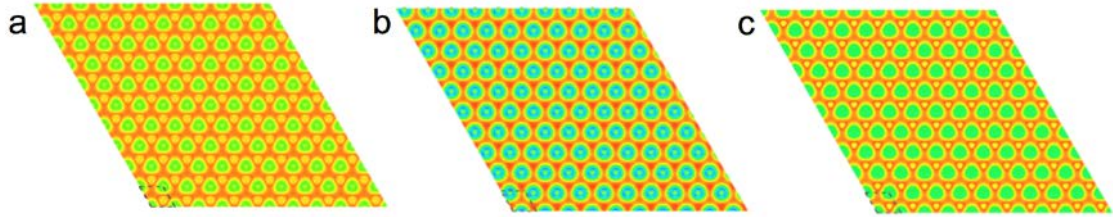

**Supplementary Figure S9. Simulated STM images of different stacking orders. (a) AB-0. (b) AB-V4. (c) AB-V3.** The bias voltage for imaging the valence band was chosen in the simulation. The images show that different configurations do not generate significant change in STM topology images, which is, most likely, because the states of top surface Se atoms dominate the appearance of topographic images. These results confirm the difficulty of distinguishing domains with different stacking orders in STM measurements. These IDB defects, showing strong metallic feature, significantly affect the contrast of STM images, making it difficult for one to pick up the minor variations (if any) of STM contrasts for different defect-free stacking domains..

**Supplementary Table S1**

The relative total energy $\Delta E_0$ (meV, with respect to the most stable stacking order AB-0), relaxation stability (whether the stacking order can stably hold their initial stacking position after relaxation), statistic number of stacking orders found in experimental ADF images by the comparison of experiments and simulations, the position of VBM ($\Gamma$ or K), interband transitions $C_Q$-$V_K$ and $C_Q$-$V_\Gamma$ (eV), the interlayer distance $d$-$Se_1$-$Se_2$ (Å). Figure 2g-h and 4f are based on the data below.

| Stacking | $\Delta E_0$ | Stability | Statistics | VBM | $C_Q$-$V_K$ | $C_Q$-$V_\Gamma$ | $d$-$Se_1$-$Se_2$ |
|---|---|---|---|---|---|---|---|
| AB-0 | 0 | y | 11 | G | 1.28 | 1.24 | 3.20 |
| AB-H1 | 44 | n | 2 | K | 1.32 | 1.40 | 3.48 |
| AB-H2 | 54 | n | 0 | K | 1.34 | 1.47 | 3.62 |
| AB-V1 | 28 | y | 0 | K | 1.31 | 1.33 | 3.34 |
| AB-V2 | 26 | y | 0 | G | 1.32 | 1.31 | 3.30 |
| AB-V3 | 22 | y | 0 | G | 1.32 | 1.31 | 3.27 |
| AB-V4 | 54 | n | 7 | K | 1.34 | 1.42 | 3.52 |
| AB-V5 | 75 | y | 0 | K | 1.35 | 1.51 | 3.76 |
| AB-V6 | 65 | n | 1 | K | 1.34 | 1.47 | 3.64 |
| AB-V7 | 25 | n | 0 | K | 1.30 | 1.33 | 3.34 |
| AA-0 | 79 | y | 0 | K | 1.32 | 1.50 | 3.76 |
| AA-H1 | 37 | n | 0 | K | 1.32 | 1.37 | 3.41 |
| AA-H2 | 20 | y | 4 | K | 1.31 | 1.32 | 3.29 |
| AA-V1 | 53 | n | 2 | K | 1.33 | 1.47 | 3.34 |
| AA-V2 | 33 | n | 5 | K | 1.32 | 1.35 | 3.35 |
| AA-V3 | 3 | y | 4 | G | 1.27 | 1.21 | 3.15 |